\documentclass[12pt]{article}
\usepackage[utf8]{inputenc}
\usepackage[margin=1in]{geometry}
\usepackage{setspace}
\usepackage{color}
\usepackage{titling}
\usepackage{graphicx}
\usepackage[bf,margin=20pt,font={small}]{caption}
\usepackage{subcaption}
\usepackage{tikz}
\usetikzlibrary{math}
\usepackage{amsmath,amssymb,dsfont}
\makeatletter\let\over\@@over\makeatother    
\makeatletter\let\atop\@@atop\makeatother 
\usepackage{hyperref}

\newcommand\TL{\hfil$\displaystyle{##}$}
\newcommand\TR{$\displaystyle{{}##}$\hfil}
\newcommand\TC{\hfil$\displaystyle{##}$\hfil}
\newcommand\TT{\hbox{##}}
\def\seqalign#1#2{\vcenter{\openup1\jot
  \halign{\strut #1\cr #2 \cr}}}
\def\lbldef#1#2{\expandafter\gdef\csname #1\endcsname {#2}}
\newcommand{\eqn}[3][]{\lbldef{#2}{(\ref{#2})}%
\begin{equation} \eqalign{#3} \label{#2} \end{equation}}
\def\eqalign#1{\vcenter{\openup1\jot
    \halign{\strut\span\TL & \span\TR\cr #1 \cr
   }}}
\def\eno#1{(\ref{#1})}

\renewenvironment{abstract}
 {\normalsize
  \begin{center}
   \bfseries \abstractname\vspace{-.5em}\vspace{0pt}
  \end{center}
  \list{}{
   \setlength{\leftmargin}{0in}%
   \setlength{\rightmargin}{\leftmargin}%
  }%
  \item\relax}
 {\endlist}

\title{Mixed field theory}

\author{Steven S.~Gubser, Christian Jepsen, Ziming Ji, and Brian Trundy}
\date{}

\begin{document}
\begin{titlingpage}

\setlength{\droptitle}{-70pt}
\maketitle
\begin{abstract}
We consider scalar field theory defined over a direct product of the real and $p$-adic numbers.  An adjustable dynamical scaling exponent $z$ enters into the microscopic lagrangian, so that the Gaussian theories provide a line of fixed points.  We argue that at $z=1/3$, a branch of Wilson-Fisher fixed points joins onto the line of Gaussian theories.  We compute standard critical exponents at the Wilson-Fisher fixed points in the region where they are perturbatively accessible, including a loop correction to the dynamical critical exponent.  We show that the classical propagator contains oscillatory behavior in the real direction, though the amplitude of these oscillations can be made exponentially small without fine-tuning parameters of the theory.  Similar oscillatory behavior emerges in Fourier space from two-loop corrections, though again it can be highly suppressed.  We also briefly consider compact $p$-adic extra dimensions, showing in non-linear, classical, scalar field theories that a form of consistent truncation allows us to retain only finitely many Kaluza-Klein modes in an effective theory formulated on the non-compact directions.
\end{abstract}
\vfill
November 2018
\end{titlingpage}

\tableofcontents

\section{Introduction}
\label{INTRODUCTION}

In this paper, we study $\phi^4$ theory defined over $\mathbb{R} \times \mathbb{Q}_p$, where $\mathbb{Q}_p$ denotes the $p$-adic numbers for a prime number $p$ which we fix once and for all.  The action we start with is
 \eqn{SphiFourth}{
  S &= {1 \over 2} \int_{\mathbb{R}} d\omega \int_{\mathbb{Q}_p} dk \, 
    \tilde\phi(-\omega,-k) \left( \omega^2 + |k|_p^{2z} + r \right) 
      \tilde\phi(\omega,k)
     + {\lambda \over 4!} \int_{\mathbb{R}} d\tau \int_{\mathbb{Q}_p} dx \, \phi(\tau,x)^4 \,.
 }
The field $\phi(\tau,x)$ is real-valued.  The parameter $z$ is real-valued as well and is understood to be a dynamical scaling exponent.\footnote{An alternative notation would be to write $\omega^2 + |k|_p^\sigma$ instead of $\omega^2 + |k|_p^{2z}$ in \eno{SphiFourth} and reserve $z$ for the dynamical scaling exponent of the full quantum theory.  Our approach is for $z$ to be defined at the level of the microscopic or classical theory, and when it differs from the infrared dynamical scaling exponent, we denote the latter by $z_\text{IR}$.}  We employ a Fourier transform
 \eqn{FourierDef}{
  \phi(\tau,x) &= \int_{\mathbb{R}} d\omega \int_{\mathbb{Q}_p} dk \, 
    e^{-2\pi i \omega\tau} \chi(kx) \tilde\phi(\omega,k)  \cr
  \tilde\phi(\omega,k) &= \int_{\mathbb{R}} d\tau \int_{\mathbb{Q}_p} dx \, 
    e^{2\pi i \omega\tau} \chi(-kx) \phi(\tau,x) \,,
 }
where $\chi(y) = e^{2\pi i \{ y \}}$ and $\{ y \}$ is the fractional part of $y \in \mathbb{Q}_p$, meaning the unique rational number in the interval $[0,1)$ such that $y-\{y\}$ is a $p$-adic integer.  We will speak of real quantities like $\tau$ and $\omega$ as Archimedean, whereas $p$-adic quantities like $x$ and $k$ will be termed ultrametric.\footnote{In general, an Archimedean norm is a positive definite norm for which if $0<|a|<|b|$, then $|na|>|b|$ for some $n \in \mathbb{Z}$.  In contrast, an ultrametric norm is a positive definite norm for which $|a+b| \leq \max\{|a|,|b|\}$.}  We refer to \eno{SphiFourth} as mixed field theory because it is defined over a space which is Archimedean in one direction and ultrametric in another.  In \eno{SphiFourth}, we write the kinetic term in Fourier space because the ultrametric part is non-local in position space.

There are several obvious generalizations of \eno{SphiFourth}.  We could replace Euclidean time $\tau \in \mathbb{R}$ by a larger Archimedean space or spacetime, for example $\mathbb{R}^d$ or $\mathbb{R}^{d-1,1}$.  We could replace $\mathbb{Q}_p$ by a more general ultrametric space, for example some extension of $\mathbb{Q}_p$, or a compact space like $\mathbb{Z}_p$ (the $p$-adic integers) or $\mathbb{P}^1(\mathbb{Q}_p)$.  We could pass to the $O(N)$ model by replacing $\phi$ with a vector $\phi^i$ valued in $\mathbb{R}^N$, and then specify a particular $O(N)$ tensor structure for the $\phi^4$ interaction.  We could consider including higher powers of $\phi$ and tuning couplings so as to arrive at multi-critical points.  Many aspects of our analysis can be straightforwardly adapted to these generalizations, so the current work should be seen as focusing primarily on the simplest example.

Broadly, our motivations are two-fold.  First, quantum systems might in principle be realized in the lab whose behavior near a critical point could plausibly be described by the mixed field theory \eno{SphiFourth}.  We have in mind especially the schemes of \cite{Britton:2012zz,Hung:2016zz}, in which non-local couplings are set up in a chain or array of quantum spins.\footnote{We thank G.~Bentsen, E.~Davis, and M.~Schleier-Smith for ongoing discussions regarding the experimental possibilities.}  Other authors have been interested in renormalization group (RG) treatments of hierarchical versions of the transverse Ising Model; a recent work along these lines is \cite{Monthus:2015zz}.

Second, we want to entertain the possibility of $p$-adic extra dimensions.  While earlier authors, notably Volovich, have advocated a sort of covariance over all possible choice of number fields in the formulations of physical theories (see for example \cite{Arefeva:1990cy,Vladimirov:1994zz}), our approach here is instead to accept $\mathbb{R}^{3,1}$ at face value as the description of the usual dimensions of spacetime, but to suppose that there could be an extra dimension which is compact and ultrametric---for example, $\mathbb{Z}_p$ for some particular $p$.

Our findings can be separated into classical and quantum results.  On the classical side, we find:
 \begin{itemize}
  \item Setting $\lambda=0$, the tree-level propagator for $\phi$ has oscillatory behavior in position space even when the argument is purely Archimedean.  For $r=0$ and $p$ not too large, this oscillatory behavior enters as a small correction to an overall power law, suppressed by a factor $e^{-{\pi^2 \over z \log p}}$.  This factor can be extremely small without fine-tuning parameters of the theory: We give an example where the suppression of oscillatory effects is by $16$ orders of magnitude.  All this is set forth in section~\ref{PROPAGATOR}.
  \item If the ultrametric domain $\mathbb{Q}_p$ is reduced to the compact set $\mathbb{Z}_p$, then, as we show in section~\ref{TRUNCATION}, there is a hierarchy of consistent truncations to purely Archimedean effective theories, based on retaining modes whose ultrametric momenta are bounded by some cutoff.  The result holds not only for $\lambda \neq 0$, but for any smooth potential $V(\phi)$, assuming $\phi=0$ is a solution to the full classical field equations.
 \end{itemize}
On the quantum side, we find:
 \begin{itemize}
  \item By power counting, as explained in section~\ref{POWER}, the $\phi^4$ interaction is irrelevant for $z<1/3$ and relevant for $z>1/3$.  We concentrate our analysis on the critical (i.e.~massless) theory close to $z=1/3$.
  \item There is evidence from one-loop computations of a Wilson-Fisher fixed point of the renormalization group when $z$ is slightly larger than $1/3$.  We present the one-loop analysis of the Wilson-Fisher fixed point in section~\ref{FIXED} after an analysis of loop diagrams in section~\ref{INTEGRALS}.  Section~\ref{MASS} exhibits a computation of the anomalous dimension of $r$ at the fixed point.
  \item Based on a two-loop computation, presented in section~\ref{UNDERGROUND}, we argue in section~\ref{DYNAMICAL} that $z$ itself is renormalized: That is, starting with \eno{SphiFourth} at the microscopic level, with $z$ slightly larger than $1/3$, the scaling parameter which enters infrared Green's functions at the Wilson-Fisher fixed point is not $|\tau|/|x|_p^z$ but rather $|\tau|/|x|_p^{z_\text{IR}}$ for a value $z_\text{IR}$ slightly different from $z$.
  \item Quantum corrections to Green's functions can have oscillatory dependence on external frequencies $\omega$, as we argue in section~\ref{POSITION} for the case of a two-loop correction to the two-point function with $z=1/3$.  This oscillatory behavior is suppressed by the same factor $e^{-{3\pi^2 \over \log p}}$ that is seen in the position space form of the tree-level Green's function.
 \end{itemize}

\section{Power counting}
\label{POWER}

Power counting in ultrametric theories is well understood: See for example \cite{Gubser:2017vgc}.  The key point is that when we scale $k \to pk$, the norm and the integration measure scale oppositely: $|k|_p \to {1 \over p} |k|_p$ and $dk \to {1 \over p} dk$.  We regard this scaling as a step toward the infrared.  We see from the kinetic term of \eno{SphiFourth} that we must accompany $k \to pk$ with $\omega \to {1 \over p^z} \omega$ and $r \to {1 \over p^{2z}} r$.  In general, we associate to a quantity $X$ an engineering dimension $[X]$ if upon a scaling $k \to pk$ we have $X \to p^{-[X]} X$.  Then the natural assignments that make $S$ dimensionless consistent with \eno{SphiFourth}-\eno{FourierDef} are
 \eqn{EngineeringDimensions}{\begin{tabular}{c|ccccccccccccc}
  $X$ &   $k$ &  $|k|_p$ & $dk$ & $x$ & $|x|_p$ & $dx$ & $\omega$ & $d\omega$ & $S$ & $\phi$          & $\tilde\phi$      & $r$  & $\lambda$  \cr \hline
  $[X]$ & $-1$ & $1$     & $1$  & $1$ & $-1$    & $-1$ & $z$      & $z$       & $0$ & ${1-z \over 2}$ & $-{3z+1 \over 2}$ & $2z$ & $3z-1$
 \end{tabular}}
We refer to these assignments as engineering dimensions because they describe scalings of the classical action without reference to loop corrections.  We see in particular that $\lambda$ has a positive dimension, meaning that $\phi^4$ is a relevant perturbation of the Gaussian fixed point theory, precisely when $z>1/3$---whereas $r$ is always relevant in the same sense, since we require $z>0$.

\begin{figure}
  \centerline{\begin{tikzpicture}
   \colorlet{darkgreen}{green!80!black}
   \tikzmath{
     \hscale = 5;
     \pointsize = 0.1;
     \p1 = \hscale;
     \p4 = 0.33 * \hscale;
     \p6 = 0.5 * \hscale;
    }
   \coordinate (C) at (0,0);
   \coordinate (P1) at (\p1,0);
   \coordinate (P4) at (\p4,0);
   \coordinate (P6) at (\p6,0);
   \coordinate (RP4) at (\p1,0.67*\hscale);
   \coordinate (N4) at (1.4*\p4,0.62*\p4);
   \coordinate (RP6) at (\p1,0.5*\hscale);
   \coordinate (N6) at (1.3*\p6,0.62*\p4);
   \coordinate (B1) at (1.15*\p1,0);
   \draw [line width=4pt,darkgreen] (C) -- (P4);
   \draw [->, line width=2pt] (C) -- (B1);
   \draw [fill] (C) circle [radius=\pointsize];
   \draw [fill] (P1) circle [radius=\pointsize];
   \node [below=5pt] at (C) {$z=0$};
   \node [below=5pt] at (P1) {$1$};
   \node [below=5pt] at (P4) {$1 \over 3$};
   \node [below=5pt] at (P6) {$1 \over 2$};
   \draw [fill,red] (P4) circle [radius=\pointsize];
   \draw [line width=2pt,red] (P4) arc (180:90:0.67*\hscale);
   \node [rotate=70,red] at (N4) {\small $\phi^4$ relevant};
   \draw [fill,red] (RP4) circle [radius=\pointsize];
   \draw [fill,blue] (P6) circle [radius=\pointsize];
   \draw [line width=2pt,blue] (P6) arc (180:90:0.5*\hscale);
   \node [rotate=70,blue] at (N6) {\small $\phi^6$ relevant};
   \draw [fill,blue] (RP6) circle [radius=\pointsize];
   \draw [fill] (1.2*\p6+0.2*\hscale,0.23*\hscale) circle [radius=0.5*\pointsize];
   \draw [fill] (1.2*\p6+0.25*\hscale,0.205*\hscale) circle [radius=0.5*\pointsize];
   \draw [fill] (1.2*\p6+0.3*\hscale,0.18*\hscale) circle [radius=0.5*\pointsize];
   \node [above=15pt,darkgreen] at (0.1*\hscale,0) {\small No relevant};
   \node [above=5pt,darkgreen] at (0.1*\hscale,0) {\small deformations};
   \node [above=10pt,left=-10pt] at (P1) {Gaussian};
   \node [right] at (B1) {$z$};
  \end{tikzpicture}}
  \caption{Conjectured pattern of critical points of \eno{SphiFourth}.}\label{CriticalPoints}
 \end{figure}
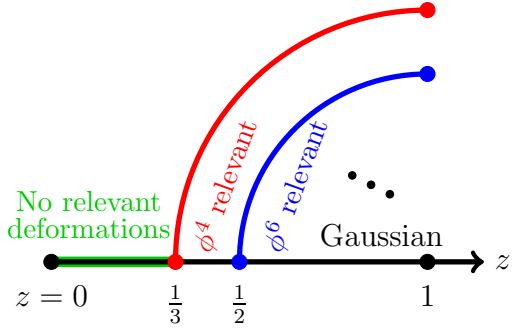

As compared to ordinary $\phi^4$ theory on $\mathbb{R}^d$, we see from the assignments \eno{EngineeringDimensions} that increasing $z$ is like decreasing $d$; that $z=1/3$ is like the upper critical dimension $d=4$, where $\phi^4$ becomes marginal; and that $z=1$ is like the lower critical dimension, where the dimension of $\phi$ goes to $0$.  Thus, at least naively, we are expecting critical points as indicated in figure~\ref{CriticalPoints}.  We added in a conjectured branch of multi-critical points based on the fact that for $z>1/2$, both $\phi^4$ and $\phi^6$ are relevant deformations of the Gaussian fixed point.  For $0<z<1/3$, our expectation based on power counting is that the Gaussian critical point is the only one available.

\section{The propagator}
\label{PROPAGATOR}

From \eno{SphiFourth} with $r=\lambda=0$, we can directly read off the momentum space Green's function:
\eqn{Gmomentum}
 {
  \tilde{G}(\omega,k)=\frac{1}{\omega^2+|k|_p^{2z}}\,.
 }
We require $\lambda=0$ for now so as to avoid loop corrections, and we require $r=0$ so as to consider the critical Gaussian theory.  Fourier transforming the real and $p$-adic components, we obtain the position space Green's function:
 \eqn{Gposition}{
  G(\tau,x) \equiv \langle \phi(\tau,x) \phi(0,0) \rangle = 
   \int_{\mathbb{R}} d\omega \int_{\mathbb{Q}_p} dk \, 
    {e^{-2\pi i\omega\tau} \chi(kx) \over \omega^2 + |k|_p^{2z}}\,.
 }
Based on the dimensional analysis of the previous section, we expect to find the scaling form
 \eqn{Gscaling}{
  G(\tau,x) = {1 \over |x|_p^{1-z}} g(\xi) \qquad\text{where}\qquad
    \xi \equiv {2\pi |\tau| \over |x|_p^z} \,.
 }
The ``aspect ratio'' $\xi$ is dimensionless in the sense $[\xi] = 0$: That is, upon scaling $x \to x/p$ and $\tau \to p^z \tau$, we find that $\xi$ is unchanged.  The factor of $2\pi$ in the definition of $\xi$ is for later convenience.  Our plan in this section is first to verify the scaling form \eno{Gscaling} and explicitly compute $g(\xi)$, and then to consider limits of small and large $\xi$.

The $\omega$ integral in \eno{Gposition} can be carried out straightforwardly, with the result
 \eqn{Ghalf}{
  G(\tau,x) = \int_{\mathbb{Q}_p} dk \, \chi(kx) {\pi \over |k|_p^z} e^{-2\pi |\tau| |k|_p^z} \,.
 }
To carry out the integral over $\mathbb{Q}_p$, we use the fact that the non-zero $p$-adic numbers split into a disjoint union of scaled copies of the $p$-adic units:
 \eqn{Qexpress}{
  \mathbb{Q}_p^\times = \bigsqcup_{v \in \mathbb{Z}} p^v \mathbb{U}_p \,.
 }
Here $\mathbb{U}_p$ is the set of all $y \in \mathbb{Q}_p$ with $|y|_p = 1$.  To handle the integration over each copy of $\mathbb{U}_p$, we use a standard result in $p$-adic Fourier analysis:
 \eqn{padicStandard}{
  \int_{p^v \mathbb{U}_p} dy \, \chi(y) = \left\{ 
    \seqalign{\span\TR &\qquad if \quad \span\TR}{
    p^{-v} / \zeta_p(1) & v \geq 0  \cr
    {-1} & v = -1  \cr
    0 & v < -1
   } \right.
 }
where we define
 \eqn{zetaDef}{
  \zeta_p(s) \equiv {1 \over 1 - p^{-s}} \,.
 }
The upshot is that \eno{Ghalf} indeed reduces to the scaling form \eno{Gscaling}, with
 \eqn{gxi}{
  g(\xi) &\equiv -{\pi \over p^z} e^{-p^z \xi} + {\pi \over \zeta_p(1)} \sum_{v=0}^\infty 
     p^{v(z-1)} e^{-p^{-vz}\xi}  \cr
   &= -{\pi \over p^z} e^{-p^z \xi} + {\pi \over \zeta_p(1)} \xi^{z-1 \over z} \sum_{v=0}^\infty 
     f\left( v - {1 \over z} \log_p \xi \right) \,,
 }
where we define
 \eqn{fv}{
  f(v) \equiv p^{v(z-1)} e^{-p^{-vz}} \,.
 }

Consider the limit $\xi \ll 1$.  We call this a highly ultrametric aspect ratio because the separation between points is mostly in the $x$ direction.  In this limit, we can ignore the factors $e^{-p^z \xi}$ and $e^{-p^{-vz}\xi}$ in \eno{gxi}, and a straightforward computation leads to
 \eqn{GSmallAspect}{
  G(\tau,x) \approx {\pi/\Gamma_p(z) \over |x|_p^{1-z}}
   \qquad\text{when}\qquad {2\pi|\tau| \over |x|_p^z} \ll 1 \,,
 }
where we define
 \eqn{GammDef}{
  \Gamma_p(z) \equiv {\zeta_p(z) \over \zeta_p(1-z)}\,.
 }

Now consider the opposite limit, $\xi \gg 1$.  We call this a highly Archimedean aspect ratio.  In this limit, we can extend the sum over $v$ in \eno{gxi} to all the integers, because $f(v)$ becomes small very quickly when $v$ goes to $-\infty$.  Now we employ Poisson resummation:
 \eqn{PoissonFormula}{
  \sum_{v \in \mathbb{Z}} f\left( v - {1 \over z} \log_p \xi \right) = 
   \sum_{\tilde{v} \in \mathbb{Z}} \xi^{-{2\pi i \over z\log p} \tilde{v}} 
    \tilde{f}(\tilde{v}) \,,
 }
where we define $\tilde{f}(\tilde{v})$ as the Fourier transform of $f(v)$,
 \eqn{fFourier}{
  \tilde{f}(\tilde{v}) \equiv \int_{\mathbb{R}} dv \, e^{-2\pi i v \tilde{v}} f(v) 
    = {1 \over z \log p} \Gamma_{\rm Euler}\left( 
       {2\pi i \tilde{v} \over z \log p} + {1 \over z} - 1 \right) \,.
 }
Noting that $\xi^{-{2\pi i \over z \log p}} = |2\pi\tau|^{-{2\pi i \over z \log p}}$ because $\xi$ differs from $|\tau|$ only by a factor of $|x|_p^z = p^{-v(x)z}$ where $v(x) \in \mathbb{Z}$ is the $p$-adic valuation of $x$, we arrive at the approximation
 \eqn{GBigAspect}{
  G(\tau,x) \approx {\pi/\zeta_p(1) \over z \log p} {1 \over (2\pi|\tau|)^{{1 \over z}-1}}
    \sum_{\tilde{v} \in \mathbb{Z}} 
     (2\pi |\tau|)^{-{2\pi i \tilde{v} \over z \log p}} 
    \Gamma_{\rm Euler}\left( 
       {2\pi i \tilde{v} \over z \log p} + {1 \over z} - 1 \right)
   \quad\hbox{when}\quad {2\pi|\tau| \over |x|_p^{z}} \gg 1 \,.
 }
In the limiting case where $x=0$, the approximate equality in \eno{GBigAspect} becomes exact.  When we compare \eno{GBigAspect} with the scaling form $G(\tau,x) = |\tau|^{1 - {1 \over z}} \tilde{g}(\xi)$ expected based on the engineering dimensions \eno{EngineeringDimensions}, it seems surprising that the infinite sum in \eno{GBigAspect} should depend directly on $\tau$ with no $x$ dependence.  In fact, because the possible values of $|x|_p$ are integer powers of $p$, we can replace $2\pi|\tau|$ by $\xi$ inside the infinite sum in \eno{GBigAspect} without changing the value of any of the terms; in other words (for $x$ not exactly $0$) the form \eno{GBigAspect} does agree with the expected scaling form.

A more illuminating way to understand the occurrence of complex powers of $|\tau|$ in \eno{GBigAspect} is to note that we only have discrete scale invariance under $x \to x/p$ and $\tau \to p^z \tau$.  Absent any other considerations, this discrete scale invariance allows the two point function $G(\tau,0)$ at purely Archimedean separation to have a power law form $|\tau|^\alpha$, but $\alpha$ can only be fixed up to additions of an integer multiple of ${2\pi i \over z \log p}$.

Despite the occurrence of complex powers of $|\tau|$ in \eno{GBigAspect}, the Green's function \eno{GBigAspect} at purely Archimedean separations is often well approximated by the $\tilde{v}=0$ term.  In fact,
 \eqn{GBigSimpler}{
  G(\tau,0) = {\pi/\zeta_p(1) \over z \log p} 
    {\Gamma_{\rm Euler}\left( {1 \over z} - 1 \right) \over (2\pi|\tau|)^{{1 \over z}-1}} h(\tau)
  \qquad\text{where}\qquad
    h(\tau) = 1 + O\left( e^{-{\pi^2 \over z \log p}} \right) \,.
 }
The function $h(\tau)$ is periodic in $\log|\tau|$: Up to an overall factor, it is the infinite sum in \eno{GBigAspect}.  Related oscillatory behavior was observed in \cite{Jagoe:2018zz}.  To understand the striking suppression factor $e^{-{\pi^2 \over z \log p}}$, we note that the Stirling approximation yields
 \eqn{StirlingApprox}{
  \left| \Gamma_{\rm Euler}\left( 
       {2\pi i \tilde{v} \over z \log p} + {1 \over z} - 1 \right) \right|
    \approx
\sqrt{2\pi}\left(\frac{2\pi}{z\log p}|\tilde{v}|\right)^{\frac{1}{z}-\frac{3}{2}}
 e^{-\frac{\pi^2}{z\log p}|\tilde{v}|}
 \,.
 }
Thus each successive term in the infinite sum in \eno{GBigAspect} is suppressed by a factor on order $e^{-{\pi^2 \over z \log p}}$.  It is worth noting that a highly suppressed amplitude of oscillation of $h(\tau)$ can be arranged without fine-tuning parameters.  For example,
 \eqn{htauEx}{
  \sup_\tau |h(\tau)-1|  
   \approx \left| \Gamma_{\rm Euler}\left( 2 + {6\pi i \over \log 2} \right) \right| \approx 10^{-16}
   \qquad\text{when $z=1/3$ and $p=2$\,.}
 }
On the other hand, this same relative amplitude $\sup_\tau |h(\tau)-1|$ rises to $0.5$ when $p$ reaches $1.75 \times 10^5$ if we hold $z=1/3$.  The relative amplitude can never exceed $1$, because one can show, starting from the first line of \eno{gxi}, that $G(\tau,x)$ is always positive, provided $z \in (0,1)$.

\section{Consistent truncation}
\label{TRUNCATION}

Consistent truncation is a phenomenon in compactification of gravitational theories (see for example \cite{Cvetic:2003jy}), whereby one can find consistent, non-linear equations of motion satisfied by a finite number of non-zero Kaluza-Klein modes.  A solution to these truncated equations of motion can be lifted back up to the original, higher dimensional theory.  It has been argued that known consistent truncations arise as a result of a generalized Scherk-Schwarz mechanism \cite{Lee:2014mla}.  Here we introduce a seemingly distinct mechanism whose essential ingredient is the assumption of ultrametricity in the compact direction.  We have no proposals in this paper for how to formulate gravitational theories over a direct product of Archimedean and ultrametric directions, so we will work instead with non-linear scalar field theory throughout.

As a warmup, consider first the action for a classical, real-valued scalar field $\phi$ defined over $\mathbb{R} \times S^1$:
 \eqn{phiFourth}{
  S = \int_{\mathbb{R}} d\tau \int_{S^1} dx \, 
    \left[ -{1 \over 8\pi^2} \phi (\partial_\tau^2 + \partial_x^2) \phi + {\lambda_n \over n} \phi^n 
      \right] \,.
 }
We could entertain a more general potential $V(\phi)$ in place of ${\lambda_n \over n} \phi^n$---for example, polynomial in $\phi$, or analytic in some neighborhood of $\phi=0$.  It is always assumed however that $\phi=0$ is a solution of the classical equation of motion, which is
 \eqn{phiFourthEom}{
  {1 \over 4\pi^2} (\partial_\tau^2 + \partial_x^2) \phi = V'(\phi) \,.
 }
We could also generalize \eno{phiFourth}-\eno{phiFourthEom} by replacing Euclidean time $\tau \in \mathbb{R}$ by Minkowski space $\mathbb{R}^{3,1}$, and the arguments to follow would be essentially unaffected.  If we assume that the $S^1$ has circumference $1$ and expand
 \eqn{phiExpand}{
  \phi(\tau,x) = \sum_{k\in \mathbb{Z}} \phi_k(\tau) e^{-2\pi i k x} \,,
 }
then unless $V(\phi)$ is purely quadratic, the only way to satisfy the equations of motion \eno{phiFourthEom} while setting all but finitely many of the Kaluza-Klein modes $\phi_k(\tau)$ to $0$ is to set all of them to zero except the mode $\phi_0(\tau)$ with no dependence at all on the $x$ direction.  Then from \eno{phiFourthEom} we obtain immediately
 \eqn{SoftModeEoms}{
  {1 \over 4\pi^2} \partial_\tau^2 \phi_0^2 = V'(\phi_0) \,.
 }
In short, the only consistent truncation of \eno{phiFourthEom} when the $x$ direction is compactified on a circle is the trivial truncation \eno{SoftModeEoms}.

Now consider a real field $\phi$ defined over $\mathbb{R} \times \mathbb{Z}_p$.  Fourier transforms can be written as
 \eqn{Pontryagin}{
  \phi(\tau,x) &= \sum_{k \in \mathbb{Q}_p/\mathbb{Z}_p} \phi_k(\tau) \chi(kx)  \cr
  \phi_k(\tau) &= \int_{\mathbb{Z}_p} dx \, \chi(-kx) \phi(\tau,x) \,,
 }
where $\mathbb{Q}_p/\mathbb{Z}_p$ is understood as a quotient of additive groups.  An element $k \in \mathbb{Q}_p/\mathbb{Z}_p$ is a coset $k = k' + \mathbb{Z}_p$, and it is easy to see that $\chi(k'x)$ for $x \in \mathbb{Z}_p$ is independent of the choice of $k'$. A canonical and unique choice is to make $k'$ its own fractional part.  Thus we can understand the sum in the first line of \eno{Pontryagin} as a sum over all rational numbers in $[0,1)$ whose denominator is a power of $p$.

The action we consider for scalar field theory on $\mathbb{R} \times \mathbb{Z}_p$ is
 \eqn{RZact}{
  S = {1 \over 2} \int_{\mathbb{R}} d\tau \sum_{k \in \mathbb{Q}_p/\mathbb{Z}_p}
    \phi_{-k}(\tau) \left( -{1 \over 4\pi^2} \partial_\tau^2 + |k|_p^{2z} \right) \phi_k(\tau) + 
   \int_{\mathbb{R}} d\tau \int_{\mathbb{Z}_p} dx \, {\lambda_n \over n } \phi(\tau,x)^n \,,
 }
and the equations of motion are
 \eqn{MixedEom}{
  {\cal D} \phi(\tau,x) = \lambda_n \phi(\tau,x)^{n-1} \,,
 }
where
 \eqn{calDDef}{
  {\cal D} \equiv {1 \over 4\pi^2} \partial_\tau^2 - D^{2z}_x
 }
and we define the Vladimirov derivative $D_x^\alpha$ of a function $f(x) = \sum_{k \in \mathbb{Q}_p/\mathbb{Z}_p} \chi(kx) \tilde{f}_k$ as 
 \eqn{DalphaDef}{
  D^\alpha_x f(x) \equiv \sum_{k \in \mathbb{Q}_p/\mathbb{Z}_p} |k|_p^\alpha \chi(kx) \tilde{f}_k \,.
 }
As before, our arguments generalize readily to polynomial $V(\phi)$ provided $\phi=0$ is a solution of the equations of motion, and to more interesting non-compact Archimedean spaces like $\mathbb{R}^{3,1}$.

The claim is that we may consistently truncate the classical equation of motion \eno{MixedEom} to soft modes:
 \eqn{Dtruncated}{
  {\cal D} \phi_s(\tau,x) = \lambda_n \phi_s(\tau,x)^{n-1} \,,
 }
where $\phi_s$ involves ultrametric Fourier modes whose norms are bounded by any cutoff we please.  Explicitly, we take a cutoff
 \eqn{LambdaM}{
  \Lambda = p^M \,,
 }
and then the requirement we place on $\phi_s$ is that it is a sum of soft modes, which are finite in number:
 \eqn{phiSoft}{
  \phi_s(\tau,x) = \sum_{k \in \mathbb{Q}_p/\mathbb{Z}_p \atop |k|_p \leq \Lambda}
    \phi_k(\tau) \chi(kx) \,.
 }
The truncated equation of motion \eno{Dtruncated} can be written without reference to $x$ as a set of coupled non-linear equations for the finite number of Kaluza-Klein modes $\phi_k(\tau)$ that we kept.  The Kaluza-Klein mass of each mode is evidently $|k|_p^{2z}$.  The equations \eno{MixedEom} and \eno{Dtruncated} have precisely the same form, so another way to phrase our claim is that the coupled equations for the soft modes $\phi_k(\tau)$ close on themselves instead of coupling to hard modes.

Since the linear operator ${\cal D}$ maps soft modes to soft modes and hard to hard, what we have to show is that hard modes aren't sourced by the right hand side of \eno{Dtruncated}: That is, any positive power of a linear combination of soft modes is still a linear combination of soft modes.  This is easy to see by induction once we know that the product of any two soft modes is still soft.  Such a product is
 \eqn{SoftTimesSoft}{
  \chi(k_1 x) \chi(k_2 x) = \chi((k_1+k_2) x)
 }
where $|k_1|_p \leq \Lambda$ and $|k_2|_p \leq \Lambda$.  Ultrametricity then indeed guarantees that $|k_1+k_2|_p \leq \Lambda$.

A point that distinguishes this ultrametric consistent truncation from the usual story in Archimedean gravitational theories is that there is a whole hierarchy of consistent truncations, where one retains all ultrametric momentum $k$ up to any prespecified maximum norm $\Lambda$, and sets all harder modes to zero.

\section{Loop integrals}
\label{INTEGRALS}

\begin{figure}
  \centering
  \begin{subfigure}[b]{0.5\textwidth}
   \centering
   \begin{tikzpicture}
    \draw[thick] (0,0) ellipse (0.6cm and 0.6cm);
    \draw [thick,-] (-1.4,-0.6) -- (1.4,-0.6);
    \draw [thick,->] (-3,-0.6) -- (-1.8,-0.6);
    \node at (-2.4,-0.3) {$\omega,k$};
    \node at (2.6,0) {$I^{(1)}_2(r,\Lambda)$};
   \end{tikzpicture}
  \end{subfigure}%
  \begin{subfigure}[b]{0.5\textwidth}
   \centering
   \begin{tikzpicture}
    \draw[thick] (0,0) ellipse (0.6cm and 0.6cm);
    \draw [thick,-] (-1.4,0.6) -- (-0.6,0) -- (-1.4,-0.6);
    \draw [thick,-] (1.4,0.6) -- (0.6,0) -- (1.4,-0.6);
    \draw [thick,->] (-3,0) -- (-1.8,0);
    \node at (-2.4,0.3) {$\omega,k$};
    \node at (3,0) {$I_4(\omega,k,r,\Lambda)$};
   \end{tikzpicture}
  \end{subfigure}\\[30pt]
  \begin{subfigure}[b]{0.5\textwidth}
   \centering
   \begin{tikzpicture}
    \draw[thick] (0,0) ellipse (0.6cm and 0.6cm);
    \draw [thick,-] (-1.4,0) -- (1.4,0);
    \draw [thick,->] (-3,0) -- (-1.8,0);
    \node at (-2.4,0.3) {$\omega,k$};
    \node at (3,0) {$I^{(2)}_2(\omega,k,r,\Lambda)$};
   \end{tikzpicture}
  \end{subfigure}
  \caption{Loop corrections to two- and four-point functions.  Total momentum $(\omega,k)$ flows into each diagram from the left.}\label{ThreeDiagrams}
 \end{figure}
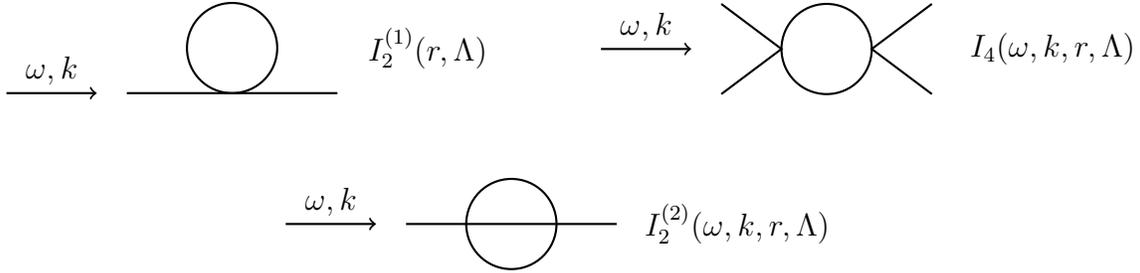

Having studied mixed field theories with compact and non-compact $p$-adic dimensions at tree-level in the preceding sections, we now turn to loop corrections, restricting our treatment to $\phi^4$ theory over $\mathbb{R} \times \mathbb{Q}_p$ as specified by \eno{SphiFourth}.  Our main goal is a renormalization group analysis of the Wilson-Fisher fixed points close to $z=1/3$.  The RG transformations we will find describe how the theory changes under a discrete scale transformation $\omega \to \omega/p^z$, $k \to pk$: See \eno{lbs} and \eno{rbs}.  In preparation, we consider in this section the loop integrations involved in evaluating the three one-particle-irreducible (1PI) Feynman diagrams shown in figure~\ref{ThreeDiagrams}.
The corresponding loop integrals are
 \eqn{LoopIntegrals}{
  I^{(1)}_2(r,\Lambda) &\equiv \int d\omega_1 \int^\Lambda dk_1 \,
    {1 \over \omega_1^2 + |k_1|_p^{2/3} + r}  \cr
  I_4(\omega,k,r,\Lambda) &\equiv \int d\omega_1 d\omega_2
    \int^\Lambda dk_1 dk_2 \,
    {\delta\left(\omega-\omega_1-\omega_2\right) \delta\left(k-k_1-k_2\right) \over
      (\omega_1^2 + |k_1|_p^{2/3} + r)(\omega_2^2 + |k_2|_p^{2/3} + r)}  \cr
  I_2^{(2)}(\omega,k,r,\Lambda) &\equiv \int d\omega_1 d\omega_2 d\omega_3
    \int^\Lambda dk_1 dk_2 dk_3 \,
    {\delta\left(\omega-\sum_{i=1}^3\omega_i\right) \delta\left(k-\sum_{i=1}^3k_i\right) \over
      \prod_{i=1}^3 (\omega_i^2 + |k_i|_p^{2/3} + r)} \,,
 }
where the superscripted $\Lambda$ on the ultrametric integrals indicates that we require all internal momenta to satisfy $|k_i|_p \leq \Lambda$, and $\Lambda = p^M$ as in \eno{LambdaM}.  Because we have only one Archimedean direction, no divergences arise from the integrations over $\omega_i$.  Thus all the integrals in \eno{LoopIntegrals} are explicitly finite.  Because the $\omega_i$ and $k_i$ integrations are of an entirely different nature, the anisotropic cut-off does not break any symmetry that would affect the universality class of the theory, just as the same holds true for the anisotropic cut-off employed in the study of the respective renormalization of the temporal and spatial derivative terms in Lifshitz scalar theory in \cite{Iengo:2009ix,Eune:2011zw}.\footnote{We note that Lifshitz scalar theories in Archimedean spacetimes have a long history, dating back as least as far as \cite{Hornreich:1975a}.}

For purposes of a renormalization group analysis, we are interested in divergences that arise as $\Lambda$ becomes large.  As we will explain in the sections~\ref{BUBBLE}, \ref{FOUR}, and~\ref{UNDERGROUND},
 \eqn{DivergenceStructure}{\seqalign{\span\TL & \span\TR & \span\TR}{
  I^{(1)}_2 &= a_{2/3} \Lambda^{2/3} + {}&A_0 r \log \Lambda^{2/3} + 
    \text{(finite as $\Lambda \to \infty$)}
    \cr
  I_4 &= &b_0 \log \Lambda^{2/3} + \text{(finite as $\Lambda \to \infty$)}  \cr
  I^{(2)}_2 &= c_{2/3} \Lambda^{2/3} + {}&(C_0 r + C'_0 \omega^2) 
    \log \Lambda^{2/3} + \text{(finite as $\Lambda \to \infty$)} \,,
 }}
for some coefficients $a_{2/3}$, $A_0$, $b_0$, $c_{2/3}$, $C_0$, and $C'_0$, all of which are independent of $\omega$, $k$, and $r$.  For the convenience of the reader who is not interested in the details of the computation, we tabulate here the coefficients that we will need for the renormalization group analysis in section~\ref{FIXED}:
 \eqn{Coefficients}{
  \begin{tabular}{c|c|c}
   $A_0$ & $b_0$ & $C'_0$ \\ \hline & & \\[-15pt]
   $\displaystyle -{3\pi \over 4\zeta_p(1) \log p}$ &
    $\displaystyle {3\pi \over 4\zeta_p(1) \log p}$ & 
    $\displaystyle 
     -{9\pi^2 \over 2 \zeta_p(1) \log p} \left( {1 - {2 \over p} \over 81} + {1 \over 8 \zeta_p(1)}
     \sum_{v=1}^\infty {p^{-2v/3} \over \left( 1 + {1 \over 2} p^{-v/3} \right)^3} \right)$
  \end{tabular}
 }
The other coefficients in \eno{DivergenceStructure} can be computed by means similar to the ones we will exhibit.

\subsection{The bubble diagram}
\label{BUBBLE}

Integrating first over $\omega_1$, we obtain
 \eqn{Isimple}{
  I^{(1)}_2 &= \pi \int^\Lambda dk_1 \, {1 \over \sqrt{|k_1|_p^{2/3} + r}}
   = \pi \int_{r^{3/2}}^\Lambda dk_1 \, \left[ {1 \over |k_1|_p^{1/3}} - {r \over 2 |k_1|_p} \right] + 
     \text{(finite)}  \cr
   &= \pi {\zeta_p(2/3) \over \zeta_p(1)} \Lambda^{2/3} - {3\pi \over 4 \zeta_p(1) \log p} r
       \log {\Lambda^{2/3} \over r} + \text{(finite)} \,,
 }
where $\int_{r^{3/2}}^\Lambda dk_1$ means that the range of integration is $r^{3/2} \leq |k_1|_p \leq \Lambda$, and we use ${}+\text{(finite)}$ to denote any dropped terms which remain finite as $\Lambda \to \infty$.  In the second equality of \eno{Isimple} we evaluate the integrals by splitting the integration domain into momentum shells; then the integrals become geometric series.  Thus we have verified the value for $A_0$ quoted in \eno{Coefficients} and checked along the way that the divergence structure of $I^{(1)}_2$ is as claimed in \eno{DivergenceStructure}.

\subsection{The four-point diagram}
\label{FOUR}

Power counting using the rules of section~\ref{POWER} allows us to check that the leading divergence in $I_4$ is logarithmic in $\Lambda$, as claimed in \eno{DivergenceStructure}.  The coefficient $b_0$ cannot depend on $\omega^2$ or $r$ because when we expand the integrand of $I_4$ in a power series in these quantities, all terms but the first (i.e.~the one that is independent of $\omega^2$ and $r$) lead to convergent integrals.  A power series in $k$ doesn't make sense for a real-valued function, but a much more powerful argument is available from ultrametricity: After a $u$-substitution $k_1 \to k_1 - k$, we can see that the integrand for $I_4$ agrees perfectly with its $k=0$ form except when $|k_1|_p \leq |k|_p$.  This immediately implies that any ultraviolet divergences are independent of $k$ and is an example of the general phenomenon of ultrametric non-renormalization; c.f.~\cite{Lerner:1989ty}.

Because $b_0$ is independent of $\omega$, $k$, and $r$, we can set $\omega=k=r=0$ in the integrand.  As before, it is useful to restrict the region of ultrametric integration to $r^{3/2} \leq |k_i|_p \leq \Lambda$.  We arrive at
 \eqn{Ifour}{
  I_4 &= \int d\omega_1 \int_{r^{3/2}}^\Lambda dk_1 \, {1 \over \left(\omega_1^2 + |k_1|_p^{2/3}\right)^2}
    + \text{(finite)}
   = {\pi \over 2} \int_{r^{2/3}}^\Lambda {dk_1 \over |k_1|_p} + \text{(finite)}  \cr
   &= {3\pi \over 4 \zeta_p(1) \log p} \log {\Lambda^{2/3} \over r} + \text{(finite)} \,,
 }
thus checking the value for $b_0$ quoted in \eno{Coefficients} as well as the overall divergence structure of $I_4$ claimed in \eno{DivergenceStructure}.

\subsection{The underground diagram}
\label{UNDERGROUND}

Power counting allows us to check that the leading divergence in $I^{(2)}_2$ scales as $\Lambda^{2/3}$, as claimed in \eno{DivergenceStructure}.  After a $u$-substitution $k_1 \to k_1 - k$, we can see that the integrand for $I^{(2)}_2$ agrees perfectly with its $k=0$ form except when $|k_1|_p \leq |k|_p$.  If we restrict the domain of integration of $k_1$ to $|k_1|_p \leq |k|_p$, then power counting shows that the integral converges.  Thus there can be no $k$ dependence in any ultraviolet divergence of $I^{(2)}_2$, and we can hereafter set $k=0$.  Expanding the integrand in a power series in $\omega^2$ and $r$ leads to the three divergent terms shown in \eno{DivergenceStructure}, followed by convergent integrals.  We are interested only in the term $C'_0 \omega^2 \log\Lambda^{2/3}$.  Therefore we can set $r=0$ in the integrand, while still restricting the region of ultrametric integration to $r^{3/2} \leq |k_i|_p \leq \Lambda$.  But we cannot set $\omega=0$ at this stage.  We introduce Schwinger parameters $u_i$, $i \in \{1,2,3\}$ in order to write
\eqn{xuDefs}{
  {1 \over \omega_i^2 + |k_i|_p^{2/3}} = \int_0^\infty du_i \, 
    e^{-u_i (\omega_i^2 + |k_i|_p^{2/3})} \,,
 }
and then the $\omega_i$ integrations in $I^{(2)}_2$ can be performed straightforwardly in terms of Gaussian integrals.  Subsequently passing to Feynman parameters $x_i = u_i/U$ where $U = \sum_i u_i$, one finds that
\eqn{JtwoFeynman}{
  I^{(2)}_2 &= \pi \int_{r^{3/2}}^\Lambda dk_1 dk_2 dk_3   
     \int_0^1 dx_1 dx_2 dx_3 \, 
      {\delta\left( 1 - {\textstyle\sum_i x_i} \right) \over 
        \sqrt{x_1 x_2 x_3 \sum_i 1/x_i}}  \, {\delta\left( {\textstyle\sum_i k_i} \right) \over
     \left( {\omega^2 \over \sum_i 1/x_i} + \sum_i x_i |k_i|_p^{2/3} \right)^2}  \cr
   &\qquad{} + \text{(finite)} \,.
 }
In order to isolate the term $C'_0 \omega^2 \log\Lambda^{2/3}$, we consider the derivative
\eqn{dJtwo}{
  {\partial I^{(2)}_2 \over \partial\omega^2} &= 
    C'_0 \log \Lambda^{2/3} + \text{(finite)} \cr
   &= -2\pi \int_{r^{3/2}}^\Lambda dk_1 dk_2 dk_3 \, 
     \int_0^1 dx_1 dx_2 dx_3 \, 
      {\delta\left( 1 - {\textstyle\sum_i x_i} \right) \over 
        \sqrt{x_1 x_2 x_3 \left( \sum_i 1/x_i \right)^3}}  \, 
        {\delta\left( {\textstyle\sum_i k_i} \right)  \over
     \left( \sum_i x_i |k_i|_p^{2/3} \right)^3}  \cr
   &\qquad\qquad{} + \text{(finite)}  \cr
   &= -\pi^2
    \int_{r^{3/2}}^\Lambda dk_1 dk_2 dk_3 \,
      {\delta\left({\textstyle\sum_i k_i}\right) \over
      |k_1 k_2 k_3|_p^{1/3} \left( {\textstyle\sum_i |k_i|_p^{1/3}} \right)^3} + 
    \hbox{(finite)}
 }
where in the third equality we used the identity
\eqn{FeynmanIdent}{
  \int_0^1 dx_1 dx_2 dx_3 \, {\delta\left( 1 - {\textstyle\sum_i x_i} \right)
     \over \sqrt{x_1 x_2 x_3 \left( \sum_i 1 / x_i \right)^3}}
    {1 \over \left( \sum_i a_i x_i \right)^3} = 
   {\pi/2 \over \sqrt{a_1 a_2 a_3} (\sqrt{a_1} + \sqrt{a_2} + \sqrt{a_3})^3} \,,
 }
valid for positive $a_1$, $a_2$, and $a_3$.

The ultrametric integral remaining in the last line of \eno{dJtwo} is immediately seen to be logarithmically divergent; indeed, we can integrate over a common scale of the $k_i$ to obtain
 \eqn{dJLog}{
  {\partial I^{(2)}_2 \over \partial\omega^2} =
   -{3\pi^2 \over 2 \log p} \log {\Lambda^{2/3} \over r} (J_\text{eq} + 3 J_\text{tall}) + 
     \text{(finite)} \,,
 }
where
 \eqn{Jeqtall}{
  J_\text{eq} &= {1 \over 27} \int_{\mathbb{U}_p} dk_1 dk_2 dk_3 \,
      \delta\left({\textstyle\sum_i k_i}\right)  \cr
  J_\text{tall} &= {1 \over 8} \int_{p\mathbb{Z}_p} dk_1 \int_{\mathbb{U}_p} dk_2 dk_3 \,
      {\delta\left({\textstyle\sum_i k_i}\right) \over
      |k_1|_p^{1/3} \left( 1+{1 \over 2}|k_1|_p^{1/3} \right)^3} \,.
 }
To reach the form indicated on the right hand side of \eno{dJLog}, we first note that the condition $k_1+k_2+k_3 = 0$ is the statement that the $k_i$, thought of as vectors in $\mathbb{Q}_p$, form the (directed) sides of a triangle.  A well-known property of ultrametric norms is that all triangles are either equilateral or tall isosceles, meaning that the base is shorter than the two equal sides.  That is, either $|k_1|_p=|k_2|_p=|k_3|_p$, or else $|k_1|_p < |k_2|_p = |k_3|_p$ up to relabeling of the $k_i$.  The integration region where the equilateral condition holds is disjoint from the three isomorphic tall isosceles regions, and each of them is disjoint from the others.  So we can carry out the full integration in the last line of \eno{dJtwo} by adding up the equilateral contribution plus three times the tall isosceles contribution with $|k_1|_p < |k_2|_p = |k_3|_p$, as in \eno{dJLog}.

To compute $J_\text{em}$, we set
 \eqn{kSet}{
  k_i = a_i + p n_i \qquad\text{where $a_i \in \mathbb{F}_p^\times$ and $n_i \in \mathbb{Z}_p$\,.}
 }
Here we identify the finite field $\mathbb{F}_p$ with the set $\{0,1,2,\dots,p-1\}$, and $\mathbb{F}_p^\times$ comprises the non-zero elements of this set.  We use the replacements
 \eqn{kintRule}{
  \int_{\mathbb{U}_p} dk_i \to 
   {1 \over p} \sum_{a_i \in \mathbb{F}_p^\times} \int_{\mathbb{Z}_p} dn_i \qquad\qquad
  \delta(k_1+k_2+k_3) \to p \delta_{a_1+a_2+a_3} \delta(n_1+n_2+n_3) \,.
 }
Using \eno{kintRule} on the expression \eno{Jeqtall} for $J_\text{eq}$, we see immediately that
 \eqn{JeqEval}{
  J_\text{eq} = {1 \over 27 p^2} \sum_{a_1,a_2,a_3 \in \mathbb{F}_p^\times} \delta_{a_1+a_2+a_3}
    = {1 \over 27\zeta_p(1)} \left( 1 - {2 \over p} \right) \,.
 }
Using the same parametrization \eno{kSet} for $k_2$ and $k_3$ in the leads to the following evaluation of $J_\text{tall}$:
 \eqn{JtallEval}{
  J_\text{tall} = {1 \over 8p} \sum_{a_2,a_3 \in \mathbb{F}_p^\times} \delta_{a_2+a_3} 
    \int_{p\mathbb{Z}_p} {dk_1 \over |k_1|^{1/3} \left( 1 + {1 \over 2} |k_1|^{1/3} \right)^3}
   = {1 \over 8 \zeta_p(1)^2} \sum_{v=1}^\infty {p^{-2v/3} \over 
        \left( 1 + {1 \over 2} p^{-v/3} \right)^3} \,.
 }
Putting \eno{dJtwo}, \eno{dJLog}, \eno{JeqEval}, and \eno{JtallEval} together, we arrive at
 \eqn{CpZero}{
  C'_0 = -{9\pi^2 \over 2 \zeta_p(1) \log p} \left( {1 - {2 \over p} \over 81} + {1 \over 8 \zeta_p(1)}
    \sum_{v=1}^\infty {p^{-2v/3} \over \left( 1 + {1 \over 2} p^{-v/3} \right)^3} \right) \,,
 }
which completes our verification of the coefficients tabulated in \eno{Coefficients}.  It may be noted that the sum over $v$ in \eno{CpZero} can be performed explicitly in terms of $q$-polygamma functions, defined as
 \eqn{qPolygamma}{
  \psi_q^{(n)}(z) \equiv {\partial^n \psi_q(z) \over \partial z^n} 
    \qquad\hbox{where}\qquad
   \psi_q(z) \equiv -\log(1-q) + 
     (\log q) \sum_{\ell=0}^\infty {q^{\ell+z} \over 1 - q^{\ell+z}} \,.
 }
Specifically,
 \eqn{SumExplicit}{
  \sum_{v=1}^\infty {p^{-2v/3} \over \left(1+{1 \over 2} p^{-v/3}\right)^3}
    = {6 \over \log p} - 
      {18 \psi_{p^{1/3}}^{(1)}\left( {3\log(-2) \over \log p}+1 \right) \over (\log p)^2} -
      {54\psi_{p^{1/3}}^{(2)}\left( {3\log(-2) \over \log p}+1 \right) \over (\log p)^3} \,.
 }

\section{The Wilson-Fisher fixed point}
\label{FIXED}

With the divergent parts of the simplest loop diagrams in hand, we can proceed to an RG analysis of the Wilson-Fisher fixed point for values of $z$ slightly larger than $1/3$, where the operator $\phi^4$ is just barely relevant.  We set
 \eqn{epsilonDef}{
  \epsilon = z - {1 \over 3}
 }
and work to the lowest non-trivial order in $\epsilon$.  In a Wilsonian picture, we start with a theory constrained by a cutoff $|k|_p \leq \Lambda$, and we find the coupling $\lambda_s$ of an effective theory for modes with $|k|_p \leq \Lambda/p$ by integrating out the modes with $|k|_p = \Lambda$.  More precisely, if $R_\Lambda$ is the region where all internal momenta $|k_i|_p \leq \Lambda$, then the modes we integrate out correspond to the integration region which is the setwise complement $R_\Lambda \backslash R_{\Lambda/p}$.

Standard analysis of the tree and one-loop diagrams for the four-point Green's function (identical to the textbook case of $\phi^4$ theory on $\mathbb{R}^4$) leads to
 \eqn{lambdas}{
  \lambda_s = \lambda - {3 \over 2} \lambda^2 \left[ I_4(\Lambda) - I_4(\Lambda/p) \right] 
    = \lambda - {3 \over 2} \lambda^2 \left[ I_4^\text{div}(\Lambda) - 
      I_4^\text{div}(\Lambda/p) \right] = \lambda - \lambda^2 b_0 \log p \,,
 }
where in square brackets we have isolated the result of restricting the integral for $I_4$ in \eno{LoopIntegrals} to the region $R_\Lambda \backslash R_{\Lambda/p}$.  In the second equality we have replaced $I_4$ by its divergent part $I_4^\text{div}$ on grounds that when evaluating the evolution of the coupling constant of the critical theory, it doesn't matter what $\omega$, $k$, and $r$ are, provided only that they are small compared to $\Lambda$.  Throughout, and in particular in the equation in \eno{lambdas}, we use our analysis from section~\ref{INTEGRALS} of loop integrals right at $z=1/3$ even though our final aim is to set $z=1/3+\epsilon$.  The reason this is permissible is that the critical point value of $\lambda$ turns out to be $O(\epsilon)$, and because we work only to the lowest non-trivial order in $\epsilon$, we can evaluate our loop integrals to zeroth order in $\epsilon$.

In terms of dimensionless couplings $\bar\lambda = \lambda/\Lambda^{3\epsilon}$ and $\bar\lambda_s = \lambda_s/(\Lambda/p)^{3\epsilon}$, \eno{lambdas} can be rewritten as
 \eqn{lbs}{
  \bar\lambda_s = p^{3\epsilon} \left[ \bar\lambda - \bar\lambda^2 b_0 \log p \right]
    = \bar\lambda + 3\epsilon \bar\lambda \log p - \bar\lambda^2 b_0 \log p \,,
 }
where we again discard terms that are beyond the lowest non-trivial order in $\epsilon$.  The fixed point condition is $\bar\lambda_s = \bar\lambda$, which immediately implies $\bar\lambda = \bar\lambda_*$ where
 \eqn{lambdaStar}{
  \bar\lambda_* = {3\epsilon \over b_0} = {4 \zeta_p(1) \log p \over \pi} \epsilon \,.
 }
This is the Wilson-Fisher fixed point for mixed $\phi^4$ theory.

\subsection{The anomalous dimension for the mass term}
\label{MASS}

Through a similarly standard analysis up to one loop for the two-point Green's function, one finds
 \eqn{rs}{
  r_s &= r + {\lambda \over 2} \left[ I^{(1)}_2(\Lambda) - I^{(1)}_2(\Lambda/p) \right] 
      = r + {\lambda \over 2} \left[ I^{(1),\text{div}}_2(\Lambda) - 
          I^{(1),\text{div}}_2(\Lambda/p) \right]  \cr 
   &= r + {\lambda \over 2} \left( a_{2/3} {\Lambda^{2/3} \over \zeta_p(2/3)} + 
       {2 \over 3} A_0 r \log p \right) \,,
 }
or, in terms of dimensionless variables $\bar{r} = r/\Lambda^{2z}$ and $\bar{r}_s = r/(\Lambda/p)^{2z}$,
 \eqn{rbs}{
  \bar{r}_s = p^{2z} \left[ \bar{r} + 
     {\bar\lambda \over 2} \left( {a_{2/3} \over \zeta_p(2/3)} + 
       {2 \over 3} A_0 \bar{r} \log p \right) \right]
    = p^{2z + \gamma_r} \left[ \bar{r} + {\bar\lambda \over 2} {a_{2/3} \over \zeta_p(2/3)} \right] \,.
 }
where the anomalous dimension of $r$ is
 \eqn{gammar}{
  \gamma_r = {\bar\lambda \over 3} A_0 \,,
 }
and we are again working to lowest non-trivial order in $\epsilon$, understanding that $\lambda \sim O(\epsilon)$.  When $\bar\lambda=\bar\lambda_*$, we see that
 \eqn{gammarStar}{
  \gamma_r = {A_0 \over b_0} \epsilon = -\epsilon \,.
 }
The value of $a_{2/3}$ enters into determining the value $\bar{r}_*$ of $\bar{r}$ at the Wilson-Fisher fixed point, but it doesn't actually matter for determining the anomalous dimension $\gamma_r$, because $\gamma_r$ has to do with the multiplicative scaling of a deviation of $r$ from the fixed point value $\bar{r}_*$.

A convenient shortcut to the result \eno{gammar} is to note that the one-particle irreducible (1PI) two-point function is
 \eqn{OnePI}{
  \Gamma^{(2)}(\omega,k) &= \omega^2 + |k|_p^{2z} + r + {\lambda \over 2} I^{(1)}_2(\Lambda)  \cr
    &= \omega^2 + |k|_p^{2z} + r + {\lambda \over 2} \left[ a_{2/3} \Lambda^{2/3} + 
        A_0 r \log {\Lambda^{2/3} \over r} \right] + \text{(finite)}  \cr
    &= \omega^2 + |k|_p^{2z} + r^{1 - {\lambda \over 2} A_0} + 
     {\lambda \over 2} \left[ a_{2/3} \Lambda^{2/3} + A_0 r \log \Lambda^{2/3} \right] + 
     \text{(finite)} \,.
 }
In the third equality of \eno{OnePI}, we are relying on the smallness of $\lambda$ to promote the $r \log r$ term to a modification of the power of $r$.  The second term in square brackets in the last line of \eno{OnePI} modifies the overall normalization of the mass term, and the first term in square brackets contributes to the critical value of $r$ where the infrared theory becomes massless.  By setting the dimensions of $r^{1 - {\lambda \over 2} A_0}$ and $|k|_p^{2z}$ equal, we see that
 \eqn{DeltaRconstraint}{
  \left( 1 - {\lambda \over 2} A_0 \right) \Delta_r = 2z \,,
 }
and upon setting $\Delta_r = 2z + \gamma_r$, we recover \eno{gammar} to leading order in $\epsilon$.

\subsection{A renormalized dynamical scaling exponent}
\label{DYNAMICAL}

We can use a similar shortcut to the one explained in \eno{OnePI}-\eno{DeltaRconstraint} to treat the effects of the underground diagram and derive a loop correction to the dynamical scaling exponent $z$.  Including the underground diagram, the 1PI two-point function is
 \eqn{GammaTwoGeneral}{
  \Gamma^{(2)}(\omega,k) &= \omega^2 + |k|_p^{2z} - {\lambda^2 \over 6} I^{(2)}_2(\Lambda) + 
     \text{($\omega$-independent)} + \text{(finite)} \,.
 }
There are contributions to $\Gamma^{(2)}(\omega,k)$ from other diagrams that include divergent terms, but these divergences have no dependence on $\omega^2$; a case in point is precisely the mass renormalization as treated in \eno{OnePI}.  In contrast, $I^{(2)}_2(\Lambda)$ has a divergent term with non-trivial dependence on $\omega$:
 \eqn{IttDiv}{
  I^{(2),\text{div}}_2(\Lambda) = C'_0 \omega^2 \log {\Lambda^{2/3} \over \omega^2} + \text{(finite)} \,.
 }
In \eno{IttDiv} we have artfully rendered the argument of the logarithm dimensionless with a factor of $\omega^2$.  This is the only sensible thing to do when studying the critical theory, because the mass is effectively zero at the critical point.  In short, 
 \eqn{GammaTwoCorrected}{
   \Gamma^{(2)}(\omega,k) &= \omega^2 + {\lambda^2 \over 6} C'_0 \omega^2 \log\omega^2 + |k|_p^{2z} -
     {\lambda^2 \over 6} C'_0 \omega^2 \log \Lambda^{2/3} +  
     \text{($\omega$-independent)} + \text{(finite)}  \cr
    &= \left(\omega^2\right)^{z/z_\text{IR}} + |k|_p^{2z} -
     {\lambda^2 \over 6} C'_0 \omega^2 \log \Lambda^{2/3} + 
     \text{($\omega$-independent)} + \text{(finite)} \,,
 }
where we have introduced an infrared dynamical scaling exponent
 \eqn{zIR}{
  z_\text{IR} = z \left( 1 - {\lambda^2 \over 6} C'_0 \right) \,.
 }
Evaluated at the Wilson-Fisher fixed point,
 \eqn{zIRvalue}{
  z_\text{IR} = z - {\epsilon^2 \over 2} {C_0' \over b_0^2}
   = z + 4 \epsilon^2 \zeta_p(1) \log p \left( {1 - {2 \over p} \over 81} + {1 \over 8 \zeta_p(1)}
    \sum_{v=1}^\infty {p^{-2v/3} \over \left( 1 + {1 \over 2} p^{-v/3} \right)^3} \right) \,.
 }

The renormalized dynamical exponent has many consequences.  For example, $\Delta_\omega = z_\text{IR}$ implies $\Delta_\tau = -z_\text{IR}$, where we always insist on $\Delta_{|k|_p} = -\Delta_{|x|_p} = 1$ for reference.  The field $\phi$ has infrared dimension
 \eqn{DeltaPhi}{
  \Delta_\phi = {1 + z_\text{IR} - 2z \over 2} \,,
 }
and in place of \eno{Gscaling} we must have an infrared scaling form
 \eqn{GscalingIR}{
  \hat{G}(\tau,x) = {1 \over |x|_p^{2\Delta_\phi}} \hat{g}(\hat\xi) \qquad\text{where}\qquad
    \hat\xi = {2\pi |\tau| \over |x|_p^{z_\text{IR}}} \,.
 }

\subsection{Position space treatment of the underground diagram}
\label{POSITION}

While the analysis of section~\ref{UNDERGROUND} seems to be the most efficient way to get hold of the coefficient $C'_0$ of the $\omega^2 \log {\Lambda^{2/3} \over \omega^2}$ term in $I^{(2)}_2(\Lambda)$ (for $z=1/3$), we would like to understand more completely the $\omega$ dependence of this loop integral, because it contributes directly to the full 1PI two-point function $\Gamma^{(2)}(\omega,k)$.  To be precise, what we will add in this section is a partial understanding of {\it finite} terms in $\Gamma^{(2)}(\omega,k)$ which are oscillatory functions of $|\omega|$, similar to the oscillatory behavior in the tree level Green's function $G(\tau,x)$ treated in section~\ref{PROPAGATOR}.  Throughout, we will work exactly at $z=1/3$ and $r=0$, understanding that the analysis can easily be extended to the Wilson-Fisher fixed point with $z=1/3+\epsilon$.  Because we are after finite terms, we remove the momentum cutoff $\Lambda$ and impose ad hoc regulators as needed.  Our methodology is based in position space, along the lines of \cite{Freedman:1991tk}.

We begin with the observation that the loop integral $I^{(2)}_2(\omega,k)$ (without a momentum cutoff) becomes simple in position space:
 \eqn{IttPos}{
  I^{(2)}_2(\omega,k) = \int_{\mathbb{R}} d\tau \int_{\mathbb{Q}_p} dx \, 
    e^{2\pi i \omega\tau} \chi(-kx) G(\tau,x)^3 \,,
 }
where, as in \eno{Gposition},
 \eqn{GtxMassless}{
  G(\tau,x) = \int_{\mathbb{R}} d\omega \int_{\mathbb{Q}_p} dk \,
    {e^{-2\pi i \omega\tau} \chi(kx) \over \omega^2 + |k|_p^{2z}} \,.
 }
Due to the complexity of Fourier transforms over $\mathbb{R} \times \mathbb{Q}_p$, we are unable to give a complete account of \eno{IttPos}.  However, the analysis simplifies somewhat when $k=0$, because then $\omega$ is the only dimensionful parameter present.  Then we claim
 \eqn{IttExpand}{
  I^{(2)}_2(\omega,0) = c_0 \omega^2 \log \omega^2 + \sum_{\tilde{v} \neq 0} c_{\tilde{v}} 
    |\omega|^{2 + {6\pi i \tilde{v} \over \log p}} \,,
 }
where $\tilde{v} \in \mathbb{Z}$.  (A pure $\omega^2$ term without the logarithm is present in \eno{IttExpand}, but its coefficient depends on the regularization scheme, so we do not attempt to compute it.)  The main technical goal of this section is to provide explicit expressions for the coefficients $c_{\tilde{v}}$, which are determined solely in terms of $p$ and $\tilde{v}$.  Comparing with \eno{IttDiv}, we see that $c_0 = -C'_0$, and a non-trivial check on our work in this section is that we can check this equality numerically, even though our explicit expression for $c_0$ is quite complicated compared to the one we gave in \eno{CpZero}.

Before getting down to detailed calculations, let's explain why the form \eno{IttExpand} is plausible.  Dimensional analysis shows that $[I^{(2)}_2] = 2/3$ when $z=1/3$, so powers $|\omega|^{2 + {6 \pi i \tilde{v} \over \log p}}$ are the ones that should contribute: These are precisely the powers of $|\omega|$ which scale by a factor of $p^{-2/3}$ when $\omega \to p^{-1/3} \omega$.  The logarithmic divergence of $I^{(2)}_2$ slightly invalidates this dimensional reasoning, but the result is only to modify the $\omega^2$ term to $\omega^2 \log \omega^2$ as shown in \eno{IttExpand}.

To verify the form \eno{IttExpand} and compute the coefficients $c_{\tilde{v}}$, a useful first step is to carry out the ultrametric Fourier transform at $k=0$ by defining
 \eqn{Fdef}{
  F(\tau) = \int_{\mathbb{Q}_p} dx \, G(\tau,x)^3
 }
so that
 \eqn{IttFromF}{
  I^{(2)}_2(\omega,0) = \int_{\mathbb{R}} d\tau \, e^{2\pi i \omega\tau} F(\tau) \,.
 }
The same dimensional reasoning used in the previous paragraph suffices to show that we can expand
 \eqn{FExpand}{ 
  F(\tau) = \sum_{\tilde{v} \in \mathbb{Z}} \tilde{c}_{\tilde{v}} 
    |\tau|^{-3 - {6\pi i \tilde{v} \over \log p}}
 }
for some coefficients $\tilde{v}$.  Only, this time there is no failure of dimensional reasoning because the integral in \eno{Fdef} is convergent provided $\tau \neq 0$.

We recognize $|\tau|^3 F(\tau)$ as a function which is periodic in $\log|\tau|$, with period ${1 \over 3} \log p$.  Thus \eno{FExpand} is essentially a Fourier series, and we can extract the coefficients $\tilde{c}_{\tilde{v}}$ from Fourier integrals:
 \eqn{ctildeInt}{
  \tilde{c}_{\tilde{v}} &= {3 \over \log p} \int_1^{p^{1/3}} d\tau \, 
    \tau^{2 + {6\pi i \tilde{v} \over \log p}} F(\tau)
   = {3 \over \log p} \int_1^{p^{1/3}} d\tau \,
      \tau^{2 + {6\pi i \tilde{v} \over \log p}} \int_{\mathbb{Q}_p} dx \, G(\tau,x)^3  \cr
   &= {3 \over \log p} \sum_{v \in \mathbb{Z}} \int_{p^{-v} \mathbb{U}_p} dx \int_1^{p^{1/3}} d\tau \,
     \tau^{2 + {6\pi i \tilde{v} \over \log p}} G(\tau,x)^3 \,.
 }
In the second line of \eno{ctildeInt} we divided the integral over $\mathbb{Q}_p$ into integrals over shells $p^{-v} \mathbb{U}_p$.  A key point, verified as usual by dimensional analysis starting from \eno{EngineeringDimensions}, is that $G(p^{v/3} \tau,p^{-v} x) = p^{-2v/3} G(\tau,x)$, so on replacing $x \to p^{-v} x$ and $\tau \to p^{v/3} \tau$ in the last line of \eno{ctildeInt}, the Jacobian factor $p^{4v/3}$ from the measure $dx d\tau$ just cancels against a factor of $p^{-4v/3}$ from the transformation of $\tau^{2 + {6\pi i \tilde{v} \over \log p}} G(\tau,x)^3$.  Thus
 \eqn{ctildeFull}{
  \tilde{c}_{\tilde{v}} = {3 \over \log p} \sum_{v \in \mathbb{Z}} \int_{\mathbb{U}_p} dx
    \int_{p^{-v/3}}^{p^{(1-v)/3}} d\tau \, \tau^{2 + {6\pi i \tilde{v} \over \log p}} G(\tau,x)^3
   = {3 \over \zeta_p(1) \log p} \int_0^\infty d\tau \,
     \tau^{2 + {6\pi i \tilde{v} \over \log p}} G(\tau,1)^3 \,.
 }
To obtain the second equality in \eno{ctildeFull}, we noticed that the intervals of $\tau$ integration neatly fit together to cover all of $\mathbb{R}$.  We also remembered that $G(\tau,x)$ depends on $x$ only through its $p$-adic norm $|x|_p$, so it is constant (at fixed $\tau$) as $x$ runs over $\mathbb{U}_p$.  Referring to \eno{Gscaling} and \eno{gxi}, we arrive at
 \eqn{Gexplicit}{
  G(\tau,1) = {\pi \over \zeta_p(1)} \sum_{v=-1}^\infty h_v p^{-2v/3} e^{-2\pi p^{-v/3} \tau} \,,
 }
where
 \eqn{hvDef}{
  h_v = \left\{ \seqalign{\span\TC &\qquad\span\TT}{
    1 & if $v \geq 0$  \cr
    {1 \over 1-p} & if $v = -1$\,.
   } \right.
 }
Plugging \eno{Gexplicit} into \eno{ctildeFull}, we arrive at
 \eqn{ctildeAgain}{
  \tilde{c}_{\tilde{v}} &= {3\pi^3 \over \zeta_p(1)^4 \log p}
    \sum_{v_1,v_2,v_3=-1}^\infty h_{v_1} h_{v_2} h_{v_3} p^{-2(v_1+v_2+v_3)/3}
     \int_0^\infty d\tau \, \tau^{2 + {6\pi i \tilde{v} \over \log p}}
      e^{-2\pi (p^{-v_1/3} + p^{-v_2/3} + p^{-v_3/3}) \tau}  \cr
   &= {3\pi^3 \over \zeta_p(1)^4 \log p}
    \Gamma_\text{Euler}\left( 3 + {6\pi i \tilde{v} \over \log p} \right)
    \sum_{v_1,v_2,v_3=-1}^\infty  
     {h_{v_1} h_{v_2} h_{v_3} p^{-2(v_1+v_2+v_3)/3} \over 
       \left[ 2\pi (p^{-v_1/3}+p^{-v_2/3}+p^{-v_3/3}) \right]^{3 + {6\pi i \tilde{v} \over \log p}}} \,.
 }

With the coefficients $\tilde{c}_{\tilde{v}}$ in hand, we now perform the real Fourier transform \eno{IttFromF} term by term in the expansion \eno{FExpand} in order to get \eno{IttExpand}.  We start with the general identity
 \eqn{GeneralFourier}{
  \int_{\mathbb{R}} d\tau \, e^{2\pi i \omega\tau} |\tau|^s = 
    {\Gamma_\infty(1+s) \over |\omega|^{1+s}} + \text{(regulator terms)} \,,
 }
where we have defined
 \eqn{GammaInfty}{
  \Gamma_\infty(s) = {\zeta_\infty(s) \over \zeta_\infty(1-s)} \qquad\text{and}\qquad
   \zeta_\infty(s) = \pi^{-s/2} \Gamma_\text{Euler}(s/2) \,.
 }
Because we are interested in finite parts of $I^{(2)}_2$, we will not concern ourselves with the regulator terms.\footnote{For completeness, let us give a short account of the regulator terms.  They are absent when the integral converges.  When there is a UV divergence (i.e.~a divergence at small $\tau$), the regulator terms are nonnegative integer powers of $\omega^2$ with divergent coefficients.  When there is an IR divergence (i.e.~at large $\tau$), the regulator terms are distributions with support at $\omega=0$, which may or may not have divergent coefficients.}  Expanding \eno{GeneralFourier} around $s=-3$ and discarding a regulator term that diverges as $1/(s+3)$, we find
 \eqn{LogFourier}{
  \int_{\mathbb{R}} d\tau \, {e^{2\pi i \omega\tau} \over |\tau|^3}
    = 2\pi^2 \omega^2 \log\omega^2 \,.
 }
Using \eno{GeneralFourier} and \eno{LogFourier}, we obtain immediately
 \eqn{cFromcTilde}{
  c_0 &= 2\pi^2 \tilde{c}_0  \cr
  c_{\tilde{v}} &= {c_{\tilde{v}} \over \Gamma_\infty\left( 3 + {6\pi i \tilde{v} \over \log p} \right)}
     \qquad \text{for $\tilde{v} \neq 0$}\,.
 }
The equality $c_0 = -C'_0$, remarked upon following \eno{IttExpand}, reduces to
 \eqn{SummationIdentity}{
  \sum_{v_1,v_2,v_3 = -1}^\infty {h_{v_1} h_{v_2} h_{v_3} p^{-2(v_1+v_2+v_3)/3} \over
    (p^{-v_1/3}+p^{-v_2/3}+p^{-v_3/3})^3} = 3 \zeta_p(1)^3 \left( {1 - {2 \over p} \over 81} + 
      {1 \over 8 \zeta_p(1)} \sum_{v=1}^\infty {p^{-2v/3} \over \left( 1 + {1 \over 2} p^{-v/3}
       \right)^3} \right) \,.
 }
It is easy to check numerically that \eno{SummationIdentity} is satisfied to good accuracy for a specified $p$, and we did so for the first $10$ primes.  But it has nothing particularly to do with prime numbers: Both sides of \eno{SummationIdentity} are easily seen to be power series in $p^{-1/3}$ with rational coefficients, and we checked that the first $20$ terms agree.  We do not have a general proof for \eno{SummationIdentity}.

The oscillatory terms $c_{\tilde{v}} |\omega|^{2 + {6\pi i \tilde{v} \over \log p}}$ in the expansion \eno{IttExpand} for $I^{(2)}_2(\omega,0)$ are quite small for modest values of $p$.  This is similar to the situation we encountered in section~\ref{PROPAGATOR} for the free propagator $G(\tau,0)$ at purely Archimedean separations.  The main source of suppression of the coefficients $c_{\tilde{v}}$ is the Euler gamma function that comes out of the integral performed in the second equality in \eno{ctildeAgain}.  A simple estimate for modest values of $p$ is
 \eqn{ctEstimate}{
  {c_{\tilde{v}} \over c_0} \sim {\Gamma_{\rm Euler}\left( 3 + {6\pi i \tilde{v} \over \log p} \right)
     \over \Gamma_\infty\left( 3 + {6\pi i \tilde{v} \over \log p} \right)} \sim 
    \sec\left( {3\pi^2 i \tilde{v} \over \log p} \right) \sim e^{-{3\pi^2 \over \log p} |\tilde{v}|} \,.
 }
We recognize in the last expression the same factor that we found in \eno{StirlingApprox} using the Stirling approximation.

\section{Conclusions}
\label{CONCLUSIONS}

In some ways, our results on $\phi^4$ theory defined over $\mathbb{R} \times \mathbb{Q}_p$ are simply what one would expect for any continuum field theory in which the dynamics is anisotropic between time and space.  Familiar features include: Power-counting as in \eno{EngineeringDimensions}, the scaling form \eno{Gscaling} for the Green's function in terms of a dimensionless variable $\xi \sim |\tau|/|x|_p^{z}$, and the use of the familiar one- and two-loop diagrams shown in figure~\ref{ThreeDiagrams} to argue perturbatively for the existence of a Wilson-Fisher fixed point when the interaction $\phi^4$ is slightly relevant.

An important difference between $\mathbb{R}$ and $\mathbb{Q}_p$ is that there is no local notion of a derivative on $\mathbb{Q}_p$; more precisely, one cannot sensibly differentiate a function mapping $\mathbb{Q}_p$ to the reals.  The closest one can come is the Vladimirov derivative, which in Fourier space amounts to multiplying by $|k|_p^s$ for some continuously adjustable real (or even complex) parameter $s$.  In the context of mixed field theory, the kernel $\omega^2 + |k|_p^{2z}$ of the kinetic term in the classical action gives immediate meaning to $z$ as the dynamical critical exponent of the microscopic theory and shows that it is an adjustable parameter.  The adjustable exponent is similar to what one sees in Archimedean theories with long-range interactions, as in \cite{Fisher:1972zz,Sak:1973zz,Sak:1977zz}; however, as far as we know, there is no analog in mixed field theory of the long-range to short-range crossover observed in theories over $\mathbb{R}^d$, as discussed recently in \cite{Behan:2017dwr,Behan:2017emf}.  Rather than changing the dimension $d$ of space, in our analysis it is natural to slide the parameter $z$ between $0$ and $1$, with $z=1/3$ playing the role of an upper critical dimension where a branch of Wilson-Fisher fixed points joins onto the line of Gaussian theories parametrized by $z$.  As we head toward $z=1$ (the analog of a lower critical dimension), more and more interaction terms $\phi^{2n}$ become relevant as deformations of the massless Gaussian theory.  (Just as in ordinary $\phi^4$ theory we can exclude odd powers of $\phi$ by imposing a $\phi \to -\phi$ symmetry.)  The overall pattern of critical points is as indicated in figure~\ref{CriticalPoints}.  Because of ultrametric non-renormalization, loop corrections to $z$ at Wilson-Fisher fixed points manifest as shifts in the power of $\omega$ entering into the 1PI two-point function: $\Gamma^{(2)}(\omega,k) \sim \left(\omega^2\right)^{z/z_\text{IR}} + |k|^{2z}_p$, where $z_\text{IR}$ is the infrared dynamical scaling exponent.

The surprising features of field theory over $\mathbb{R} \times \mathbb{Q}_p$ are connected to the fact that Archimedean norms like $|\tau|$ and $|\omega|$ are continuously variable, whereas the $p$-adic norms $|x|_p$ and $|k|_p$ take discrete values: integer powers of the chosen prime $p$.  The most scale invariance one can hope for is invariance under discrete transformations $x \to x/p$ and $\tau \to p^z \tau$.\footnote{If the infrared theory has a renormalized dynamical critical exponent $z_\text{IR} \neq z$, then the appropriate scaling in the infrared is $x \to x/p$ while $\tau \to p^{z_\text{IR}} \tau$.}  A result of this weakened scale invariance is that we can have oscillatory dependence of Green's functions on Archimedean quantities.  For example, when $z=1/3$, we find by dimensional analysis that the two-point Green's function at purely Archimedean separation is $G(\tau,0) \sim 1/\tau^2$.  But all that our discrete scale invariance really implies is that $\tau^2 G(\tau,0)$ is periodic in $\log |\tau|$, with period ${1 \over 3} \log p$.  In other words, $\tau^2 G(\tau,0)$ is a sum of integer powers of $|\tau|^{6\pi i \over \log p}$.  Non-constant log-periodic behavior in $\tau^2 G(\tau,0)$ appears already in the massless Gaussian theory, as explained in detail in section~\ref{PROPAGATOR}.  Two-loop corrections to the 1PI two-point function $\Gamma^{(2)}(\omega,k)$ of the critical theory at $z=1/3$ involve similar log-periodic behavior in $\omega$ when $k=0$, modified by the $\omega^2 \log \omega^2$ that loop-corrects the dynamical scaling exponent.  Specifically, as explained in section~\ref{POSITION}, $\omega^{-2} \Gamma^{(2)}(\omega,0)$ includes all integer powers of $|\omega|^{6\pi i \over \log p}$: precisely the powers which are invariant under $\omega \to p^{-1/3} \omega$.

More surprising than the presence of oscillatory terms permitted by the discrete scale invariance that characterizes critical mixed field theories is the smallness of their coefficients.  In both the classical and two-loop examples discussed in the previous paragraph, we found by explicit calculation that oscillatory terms are suppressed by a factor $e^{-{3\pi^2 \over \log p}}$.  This factor is extremely small for modest values of $p$.  For general values of $z$, the analogous factor is $e^{-{\pi^2 \over z \log p}}$.  We would like to reach an understanding of how such a factor could emerge from an instanton construction, hints of which might already be found in our use of Poisson summation in \eno{PoissonFormula}.

Many generalizations of our constructions could be considered, as enumerated already in section~\ref{INTRODUCTION}.  Passing to the $O(N)$ model is straightforward, and at least at the level of fixed orders in perturbation theory, all our arguments go through unaltered, with appropriate functions of $N$ multiplying the contributions of the various loop diagrams.  A large $N$ Hubbard-Stratonovich treatment of the critical $O(N)$ model appears much more challenging, because the requisite Fourier transforms, which lead to simple powers in the purely Archimedean and purely ultrametric cases \cite{Gubser:2017vgc}, instead lead in mixed field theory to complicated scaling forms similar to the ones considered in sections~\ref{PROPAGATOR} and~\ref{POSITION}.

Passing to a finite extension of $\mathbb{Q}_p$ should be straightforward.  For unramified extensions, the main changes would be the replacement $p \to p^n$ and some related alterations in power counting.  For ramified extensions, the norm of the uniformizer would affect how coarse the scaling symmetry is in the Archimedean direction: For example, if $\sqrt{p}$ is adjoined to $\mathbb{Q}_p$ and the action still contains $\omega^2 + |k|_p^{2z}$, then $\omega \to p^{-z/2} \omega$ as $k \to \sqrt{p} k$.

Our original phenomenological motivation suggests generalizing Archimedean Euclidean time $\tau \in \mathbb{R}$ to a four-vector in $\mathbb{R}^{3,1}$.  A new feature would then be divergences already in the Archimedean parts of loop integrals.  A more careful treatment of how momentum cutoffs can be applied in Archimedean and/or ultrametric directions would be called for, hopefully with the result that there is not much sensitivity to details of the cutoff in the scaling properties of the infrared effective theories.  Compact ultrametric dimensions are obviously another interesting avenue to explore further, with many mysteries at the outset.  Massless fields with spin might be treated in terms of appropriate multiplicative characters, as in \cite{Gubser:2018cha}, but gauge fields seem particularly difficult given the totally disconnected topology of $\mathbb{Q}_p$.

\section*{Acknowledgments}

This work was supported in part by the Department of Energy under Grant No.~DE-FG02-91ER40671, and by the Simons Foundation, Grant 511167 (SSG).

\clearpage

\bibliographystyle{utphys}
\bibliography{mixed}

\end{document}